\documentclass{ws-b8-5x6-0}
 \usepackage{amssymb,amsmath}  % AmsTeX Packages
 \usepackage{graphicx}

\DeclareMathOperator{\Tr}{Tr}

\usepackage{graphicx,latexsym}
\usepackage{amssymb,amsmath}  % AmsTeX Packages
\usepackage{amsfonts}         % AmsTeX Packages
\usepackage{dcolumn}          % Align table columns on decimal point
\usepackage{bm}               % bold math
\usepackage{graphicx}

\newcommand{\gbTr}{\mbox{Tr}}

\begin{document}
 
\chapter{Two Quark Potentials} 
\author{G. Bali} 
\addauthors{G. Bali} 
 
\address{ Department of Physics \& Astronomy, University of Glasgow,\\
 Glasgow G12 8QQ, Scotland\\
E-mail: g.bali@physics.gla.ac.uk} 
\renewcommand{\theequation}{\arabic{chapter}.\arabic{equation}}

\begin{abstract}
In this Chapter QCD interactions between a quark and an anti-quark are 
discussed.
In the heavy quark limit these potentials can be related to quarkonia
and $1/m$ corrections can be systematically determined. Excitations
of the ground state potential provide an entry point into the
phenomenology of quark-gluon hybrids. The short-distance behaviour
of non-perturbative potentials can serve as a test of
resummation and convergence of perturbative expansions.
Torelons and potentials between non-fundamental
colour charges offer a window into the origin of the confinement
mechanism and relate to effective string descriptions of low energy
aspects of QCD.
\end{abstract}
\section{Motivation}
In order to avoid excessive overlap with the many introductions that already
exist in different places of this Volume, we shall elaborate on
a very subjective motivation of studying interquark potentials in QCD,
centred around the general theme of {\em building bridges} between
models and QCD. 

QCD contains two related, non-perturbative features:
the breaking of (approximate) chiral symmetry and the (effective) confinement
of coloured objects such as quarks and gluons. Similarly, models of the QCD vacuum
can roughly be divided into two classes: those that are primarily based on
chiral symmetry and those that have confinement, for instance in the form of
a confining potential, as their starting point. Often it is difficult to
{\em microscopically} relate a particular model to QCD.
For example,  should the instantons that are
evidenced in lattice simulations within a given prescription at a finite
cut-off, be the same that are supposed to appear within instanton liquid
models? The answer to this question is not known.

There exist, however,
two ways of systematically {\em building bridges}. Perhaps the most obvious
one is the comparison of {\em global} properties like form factors,
charge distributions in position space, potential energies or particle masses.
Such a comparison circumvents the problem of identifying a one-to-one
mapping of the degrees of freedom within a given model, which often
might not qualify as a quantum field theory, onto objects that appear
in the QCD vacuum.
The model would then very much resemble the pragmatic use of analogies
in the popular science literature: it is not that important to
get things 100~\% right if we cannot understand them 100~\% anyway.

A lot of purpose-engineered QCD information that is not directly accessible
to experiment can be {\em manufactured} in lattice simulations. Moreover, in
experiment there is only one world. On the lattice one has the freedom
to vary the number of sea quark flavours $n_f$, the quark masses, the number of
gauge group colours, $N_C$, the volume {\it etc.\ etc.}, away from the values that
happen to be realized in nature for one reason or another.
Sometimes (like in the case of quark masses) this is at present
a necessity, in other cases it is pure virtue. In having not only one
{\em physical world} but many {\em virtual worlds} at ones disposal, the
applicability range and precision to be expected from any model can, in
principle, be tested very stringently. Unfortunately, in practice,
such interaction between lattice practitioners and model builders is
still rather under-developed for various psychological, communication-related
and dogmatic reasons and, in some cases, even out of fear, mistrust or
over-confidence, one might speculate.

The other way of {\em building bridges} is when a separation of scales
occurs, in which the symmetries of QCD constrain the number of possible
terms and allow for the systematic construction of an effective
field theory. The most prominent examples are the chiral effective field
theory ($\chi$EFT) that governs the low energy interactions of QCD
as well as heavy quark effective theory (HQET) and non-relativistic
QCD (NRQCD) for heavy quark physics. While in $\chi$EFT the scale separation
occurs through the spontaneous breaking of the chiral symmetry by
some collective gluonic effect, in HQET and NRQCD
the scale separation is provided by
the heavy quark mass $m$. Ideally,  one would calculate the
``high'' energy Wilson coefficients of $\chi$EFT,
such as the pion decay constant $F_{\pi}$, in lattice simulations (or 
determine them from fits to experimental data).
The low energy expansion, as a function of $m_{\pi}$, can be determined
analytically (chiral perturbation theory). On the other hand, in
HQET and NRQCD the low energy matrix elements can be provided by
non-perturbative lattice simulations while the Wilson coefficients are
calculable in perturbation theory, as long as $m\gg\Lambda$, where
$\Lambda$ denotes a typical non-perturbative scale of order 400~MeV.

It is this latter heavy quark limit, in which the static QCD potentials
that are discussed here --- as well as in Chapter~4 --- can be related to
mesonic bound states and, in the case of three body potentials, baryonic bound states.
In doing this, QCD can be reduced to non-relativistic quantum mechanics within
this particular sector, with the help of reasonable assumptions which
themselves can be tested in lattice simulations. For example, in the case of quarkonia,
QCD itself tells us what ``potential model'' we have to choose.

It is also possible
to fit the spectra of light mesons and baryons, assuming phenomenological
potentials. These potentials cannot be related in a systematic way
either to QCD or to static potentials as calculated from Wilson loops.
However, in spite of some rather dubious assumptions
that are implicitly folded into such
non-relativistic or relativized quark-potential
models, it can still be instructive
to compare the parametrizations that are commonly used in this context
with those that are relevant
in quarkonium physics. After all, most {\em physics} appears to interpolate
smoothly between hadrons made out of light and heavy quarks. There also
appears to be an intimate
connection between, on one hand, the  broken chiral symmetry that is most
relevant in the light hadronic sector --- but plays little direct
r\^ole for quarkonia
with masses much larger than the chiral symmetry breaking scale --- and, on the
other hand, the confining potential and flux tubes that seem to be the
vacuum excitations that are responsible for interactions between the slowly
moving heavy quark degrees of freedom.

\section{The Static QCD Potential}
\label{potential}
The very definition of a ``potential'' requires the concept of
``instantaneous'' interactions: a test particle has to interact
with the field induced by a source on a time scale short enough
to guarantee that the relative distance remains unaffected.
Whenever the relative speed of the two particles becomes
relativistic, the underlying assumption of a constant time
difference between cause and effect
is obviously violated: only in non-relativistic systems, 
{\it i.e.}\ as long as the
typical interaction energies ($E$)
within bound states are small compared to
the particle masses ($M$),  can we define a potential as a function of
coordinates such as the
distance $r$, the spin $S$, angular momentum $L$ and relative momentum $p$.
While for interactions between elementary charges in QED this is always the
case as $E/M=O(\alpha_F)\ll 1$, QCD implies typical binding
energies of order 400~MeV and only
the bottom, and eventually the charm quark, can be regarded as
non-relativistic.

Another example in which the non-relativistic approximation is justified
are nucleon-nucleon potentials, $V_{NN}(r)$. Although there exist attempts to
extract this information also from QCD, by employing lattice simulations
(cf.\ Chapter~4 Subsec.~4.4.2),
it is fair to predict that quantitatively reliable information, using
the presently available methodology,
might not become available within this decade. However, in this case a wealth
of phenomenological information exists from experiment. For the
quark-antiquark potential it is exactly the opposite. All experimental 
information is model dependent and rather indirect as quarks never appear
as free particles in nature. However, this potential is among the most
precisely determined quantities that have been calculated so far on the lattice

As a starting point one can make the test charges infinitely heavy,
prohibiting
any change in their relative speed and study the ``static'' limit.
To this end we shall introduce the Wegner-Wilson loop and derive its
relation to
the static potential. Subsequently, expectations of this potential
from exact considerations, strong coupling and string arguments
as well as from perturbation theory
are presented. Lattice results are then
reviewed. Finally, the model is extended to non-static quarks and the
form of the resultant potential in coordinate space is compared with its
counterpart ($\omega$-meson exchange) in the nucleon-nucleon interaction.

\subsection{Wilson loops}
\label{sec:will}
We will derive the relationship between the expectation values
of Wegner-Wilson loops and the
potential energy $V(r)=E_0(r)$ between two colour charges, separated by
a distance $r$. This is a technical but instructive exercise. The final
result is displayed in Eq.~\ref{eq:specw}.

\begin{figure}[thb]
%\centerline{\includegraphics[height=84mm]{wilsonpic.eps}}
\centerline{\includegraphics*[height=0.5\textwidth]{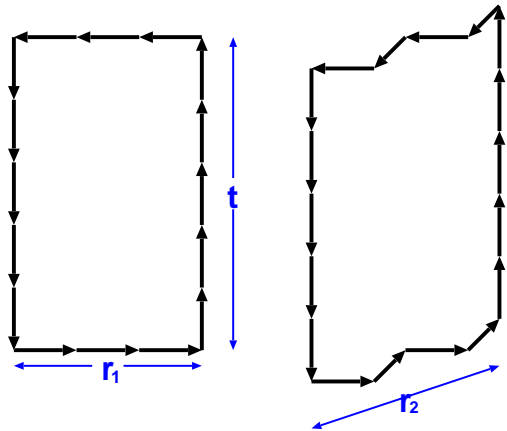}}
\vspace*{8pt}
\caption{Examples of rectangular on- and off-axis Wilson loops with temporal
extent $t=5a$ and spatial extents $r_1=3a$ and
$r_2=2\sqrt{2}\,a$, respectively.
\label{figwilson}}
\end{figure}

The Wegner-Wilson loop was originally  introduced by
Wegner~\cite{Wegner:1984qt}
as an order parameter in $Z_2$ gauge theory.
It is defined as the trace of the product of gauge variables $U_{x,\mu}$ along
a closed oriented contour $\delta C$, enclosing \mbox{an area $C$,}
\begin{equation}
\label{eq:wl}
W(C)=\Tr\left\{
{\mathcal P}\left[\exp\left(
i\int_{\delta C}\!dx_{\mu}\,A_{\mu}(x)\right)\right]\right\}=
\Tr\left(\prod_{(x,\mu)\in\delta C}U_{x,\mu}\right).
\end{equation}
While the loop, determined on a gauge configuration\footnote{
$U_{\,\mu}\in SU(3)$ a gauge group element, pointing into direction
$\mu\in\{1,2,3,4\}$, located between positions $x$ and $x+a\hat{\mu}$,
where $a$ is the lattice spacing and $x$ is a lattice point within
the (finite) 4-volume.}
 $\{U_{x,\mu}\}$,
is in general complex, its expectation value is real, due to
charge invariance: in Euclidean space we have
$\langle W(C)\rangle=\langle W^*(C)\rangle=\langle W(C)\rangle^*$.
It is straight forward to generalise the above Wilson loop
to any non-fundamental representation $D$ of the gauge field,
just by replacing
the variables $U_{x,\mu}$ with the corresponding links $U_{x,\mu}^D$.
The arguments below, relating the Wilson loop to the potential
energy of static sources, go through independent of the representation
according to which the sources transform under local
gauge transformations. In what follows, we will denote a Wilson loop,
enclosing a rectangular contour with one purely spatial distance,
${\mathbf r}$, and one temporal separation, $t$, by $W({\mathbf r},t)$.
Examples of Wilson loops on a lattice for two different choices
of contours $\delta C$ are displayed in Figure~\ref{figwilson}.

In Wilson's original work~\cite{Wilson:1974skgb},
the Wilson loop had been related to
the potential energy of a pair of static colour sources by using
transfer matrix arguments. However,
it took a few years until
Brown and Weisberger attempted to
derive the connection between the Wilson loop
and the effective
potential between heavy, not necessarily static, quarks in a mesonic
bound state~\cite{Brown:1979ya}.
Later on, mass dependent
corrections to the static potential have been
derived along similar lines~\cite{Eichten:1979pu,Barc12}
and the approach has been made systematic within the framework of
EFTs~\cite{Pineda:2000sz,Chen:1995dg,Bali:1997am,Brambilla:2000gk}.
In Sec.~\ref{sec:gluon}, 
we will discuss these developments in more detail.
Here, we
derive the connection between a Wilson loop
and the static potential between colour sources
which highlights similarities with
the situation in classical electrodynamics.

For this purpose we start from the Euclidean Yang-Mills action,
\begin{equation}
S=\frac{1}{4g^2}\int\!d^4x\,F_{\mu\nu}^aF_{\mu\nu}^a.
\end{equation}
The canonically conjugated momentum to the field, $A_{i}^a$, is given by
the functional derivative,
\begin{equation}
\label{eq:can1}
\pi^a_{i}=\frac{\delta S}{\delta(\partial_4A_{i}^a)}=\frac{1}{g^2}
F_{4i}^a=-\frac{1}{g}E_i^a.
\end{equation}
The anti-symmetry of the field strength tensor implies $\pi^a_4=0$.
In order to obtain a Hamiltonian formulation of the
gauge theory, we fix the temporal gauge {\it i.e.} $A_4^a=0$. 
In infinite volume
such gauges can always be found. On a toroidal lattice
this is possible up to one time slice $t'$, which we demand to be outside of
the Wilson loop contour, {\it i.e.} $t'>t$.

The canonically conjugated momentum,
\begin{equation}
\label{eq:can2}
\pi_{\mu}^a=-i\frac{\delta}{\delta A_{\mu}^a},
\end{equation}
now fulfils the usual commutation relations,
\begin{equation}
[A_{j}^a,\pi_{\mu}^b]=i\delta_{j \mu}\delta^{ab},
\end{equation}
and we can construct the Hamiltonian,
\begin{equation}
H=\int\!d^3x\,\left(\pi_{\mu}^a\partial_4A_{\mu}^a-\frac{1}{4g^2}
F_{\mu\nu}^aF_{\mu\nu}^a\right)
=\frac{1}{2}\int\!d^3x\,\left(E_i^aE_i^a-B_i^aB_i^a\right)
\end{equation}
that acts onto states $\Psi[A_{\mu}]$. In the Euclidean metric
the magnetic contribution to the total energy is negative.
Note that we can also add a fermionic
term $\sum_f\bar{q}_f[\gamma_{\mu}D_{\mu}+m_f]q_f$ to the action above.
In this case the momentum conjugate to the Dirac spinor field
$q_f^{\alpha}$ is given by
$-i\delta/\delta q_f^{\alpha}=\bar{q}_f^{\alpha}\gamma_4$,
where $\alpha=1,\cdots,N_C$
runs over the colour in the fundamental representation.
Here $f$ denotes the quark flavour and $m_f$ the respective mass.

A gauge transformation $\Omega$
can be represented as a bundle
of $SU(N_C)$ matrices in some representation $R$,
$\Omega_R({\mathbf x})=e^{i\omega^a({\mathbf x})T_R^a}$.
We wish to derive the
operator representation
of the group generators $T_R^a$, that acts on the Hilbert space
of wave functionals.
For this purpose
we start from the definition
\begin{equation}
R(\Omega)\Psi=
\left[1+i\int\! d^3x\, \omega^a({\mathbf x})
T_R^a({\mathbf x})
+\cdots\right]\Psi=\Psi+\delta \Psi.
\end{equation}
One easily verifies that
$\delta A_i=A_i^{\Omega}-A_i=-(\partial_i\omega+i[A_i,\omega])
\equiv D_i \omega({\mathbf x}) $.
We then obtain
\begin{eqnarray}
\delta\Psi&=&
\int\!d^3\!x\,\delta A_i({\mathbf x})
\frac{\delta\Psi}{\delta A_i({\mathbf x})}
=\int\!d^3\!x\,D_i\omega({\mathbf x})
\frac{\delta\Psi}{\delta A_i({\mathbf x})}\nonumber\\
&=&-\frac{i}{g}\int\!d^3\!x\,\omega^a({\mathbf x})(D_iE_i)^a({\mathbf x})
\Psi,
\end{eqnarray}
where we have performed a partial integration and have
made use of the equivalence
\begin{equation}
\frac{\delta}{\delta A_i}=-\frac{i}{g}E_i
\end{equation}
of Eqs.~\ref{eq:can1} and \ref{eq:can2}.
Hence we obtain the representation
\begin{equation}
T_R^a=-\frac{1}{g}(D_iE_i)^a,
\end{equation}
{\it i.e.} the covariant divergence of the electric field operator is the
generator of gauge transformations! Again note that had we included
sea quarks into the action, we would have encountered an additional term
$-\frac{1}{g}\sum_f\bar{q}_f\gamma_4T^aq_f$ on the right hand side of this equation,
where the generator is to be taken in the fundamental representation.
It is trivial to generalize the equations below accordingly and the
physical meaning is clear too: the vacuum has an intrinsic charge density
distribution due to sea quarks. 
%As we will see in Sec.~\ref{sec:string},
This then in turn
allows for {\em string breaking} of the static potential.

Let us assume that the wave functional is a singlet
under gauge transformations
$
R(\Omega)\Psi[A_{\mu}]=\Psi[A_{\mu}].
$
This implies that
\begin{equation}
(D_iE_i)^a\Psi=0,
\end{equation}
which is Gauss' law in the absence of sources:
$\Psi$ lies in the eigenspace of $D_iE_i$
that corresponds to the eigenvalue zero.
Let us next place an external source in the fundamental representation of the
colour group at position ${\mathbf r}$.
In this case, the associated wave functional
$\Psi_{\alpha},\alpha = 1,\cdots,N_C$ transforms in a non-trivial way,
namely
\begin{equation}
[R(\Omega)\Psi]_{\alpha}
=\Omega_{\alpha\beta}\Psi_{\beta}.
\end{equation}
This implies that
\begin{equation}
\label{eq:gau}
(D_iE_i)^a\Psi=-g\delta^3({\mathbf r})T^a\Psi,
\end{equation}
which again resembles Gauss' law, this time
for a point-like colour charge at position\footnote{Of course,
on a torus, such a state cannot be constructed. Note that in our Euclidean
space-time conventions Gauss' law reads $[{\mathbf D}{\mathbf E}]^a
({\mathbf x})=-\rho^a({\mathbf x})$, where $\rho$ denotes the charge density.
Again note that, in general, $\rho^a$ will automatically contain a contribution
$g\bar{q}_f\gamma_4T_aq_f$ for each sea quark flavour $f$.}
${\mathbf r}$.
For non-fundamental representations $D$,
Eq.~\ref{eq:gau} remains valid under the replacement 
$T^a\rightarrow T^a_D$.

Let us now place a fundamental
source at position ${\mathbf 0}$ and an anti-source
at position ${\mathbf r}$. The wave functional $\Psi_{\mathbf r}$,
which is an $N_C\times N_C$ matrix in colour space
will transform according to
\begin{equation}
\Psi^{\Omega}_{\mathbf r,\alpha\beta}
=\Omega_{\alpha\gamma}({\mathbf 0})\Omega^{*}_{\beta\delta}({\mathbf r})
\Psi_{\mathbf r,\gamma\delta}.
\end{equation}
One object with the correct transformation property
is a gauge transporter (Schwinger line) from ${\mathbf 0}$ to
${\mathbf r}$,
\begin{equation}
\Psi_{\mathbf r}
=\frac{1}{\sqrt{N_C}}
U^{\dagger}({\mathbf r},t)=\frac{1}{\sqrt{N_C}}
{\mathcal P}\left[\exp\left(
i\int_{\mathbf 0}^{\mathbf r} d{\mathbf x}\,{\mathbf A}({\mathbf x},t)\right)
\right],
\end{equation}
which on the lattice corresponds to the ordered product of link variables
along a spatial connection between the two points.
Since we are in the temporal gauge, $A_4(x)=0$, the correlation function
between two such lines at time-like separation $t$
is the Wilson loop
\begin{equation}
\langle W({\mathbf r},t)\rangle=\frac{1}{N_C}
\langle U_{\alpha\beta}({\mathbf r},t)
U^{\dagger}_{\beta\alpha}({\mathbf r},0)\rangle,
\end{equation}
which, being a gauge invariant object, will give the same result in any gauge.
Other choices of $\Psi_{\mathbf r}$, {\it e.g.}\ linear combinations of
spatial gauge transporters connecting ${\mathbf 0}$ with ${\mathbf r}$,
define generalised (or smeared) Wilson loops, $W_{\Psi}({\mathbf r},t)$.

We insert a complete
set of transfer matrix eigenstates, $|\Phi_{{\mathbf r},n}\rangle$,
within the sector of the Hilbert space
that corresponds to a charge and anti-charge in the fundamental
representation at distance ${\mathbf r}$, and expect the Wilson loop
in the limit of large temporal lattice extent, $L_{\tau}a\gg t$,
to behave like
\begin{equation}
\label{eq:specw}
\langle W_{\Psi}({\mathbf r},t)\rangle
=\sum_n
\left|\left\langle \Phi_{\mathbf r,n}\right|\Psi_{{\mathbf r}}\left|
0
\right\rangle
\right|^2
e^{-E_n({\mathbf r})t},
\end{equation}
where the normalisation convention
is such that $\langle\Phi_n|\Phi_n
\rangle=\langle\Psi^{\dagger}\Psi\rangle=1$ and the completeness of
eigenstates implies $\sum_n |\langle \Phi_n|\Psi|0\rangle|^2=1$.
Note that no disconnected part has to be subtracted from
the correlation function since
$\Psi_{\mathbf r}$ is distinguished from
the vacuum state by its colour indices. 
$E_n({\mathbf r})$
denote the energy levels. The ground state
contribution $E_0({\mathbf r})$ --- that will
dominate in the limit of large $t$ --- can
now be identified with the static potential $V({\mathbf r})$,
which we have been aiming to calculate.

The gauge transformation properties of the colour state discussed above,
which determine the colour group representation of the static sources
and their separation ${\mathbf r}$, do not yet completely
determine the state in question: the sources will be connected by
an elongated chromo-electric
flux tube.
This vortex can, for instance, be in a rotational state
with spin $\Lambda\neq 0$ about the inter-source
axis. Moreover, under interchange of
the ends
the state can transform evenly ($\eta=\mbox{g}$) or oddly ($\eta=\mbox{u}$),
where $\eta$ denotes the combined $CP$ parity.
Finally, in the case of the one-dimensional $\Lambda = 0$
representations, it can transform symmetrically
or anti-symmetrically
under reflections with respect to a plane containing the sources
($\sigma_v=\pm$).
It is possible to
single out sectors within a given irreducible representation of the
relevant cylindrical symmetry group~\cite{Landau:1987qm}, $D_{\infty h}$,
with an adequate choice of $\Psi$. 
A straight line connection between the sources
corresponds to the $D_{\infty h}$ quantum numbers
$\Sigma_g^+$, where $\Lambda=0,1,2,\cdots$ is replaced by capital
Greek letters, $\Sigma,\Pi,\Delta,\cdots$.
Any static potential that is different from the 
$\Sigma_g^+$ ground state
will be referred to as a ``hybrid''
potential.
\footnote{Compare the discussion in Sec.~2.3.1 of Chapter~2}
 Since these potentials are gluonic excitations
they can be thought of as being hybrids between pure ``glueballs''
and a pure static-static state;
indeed, high hybrid excitations are
unstable and will decay into lower lying potentials via the radiation of
glueballs. 
%We will address the question of hybrid potentials in detail in
%Secs.~\ref{sec:other}, \ref{sec:potnrqcd} and \ref{sec:pertr}.

\subsection{Exact results}
We identify the static potential $V({\mathbf r})$
with the ground state energy
$E_0({\mathbf r})$ of Eq.~\ref{eq:specw} that can be extracted from the
Wilson loop of Eq.~\ref{eq:wl}.
By exploiting the symmetry of a Wilson loop under an interchange of the space and
time directions, it can be proven that the static potential cannot
rise faster
than linearly as a function of the distance $r$ in the limit
\mbox{$r\rightarrow\infty$~\cite{Seiler:1978ur}}.
Moreover, reflection
positivity of Euclidean $n$-point
functions~\cite{Osterwalder:1973dx,Osterwalder:1975tc} implies
convexity of the static
potential~\cite{Bachas:1986xs}, {\it i.e.}
\begin{equation}
V''(r)\leq 0.
\end{equation}
The proof also applies to ground state
potentials between sources in
non-fundamental representations.
However, it does not apply to hybrid excitations,
since in this case the required creation operator
extends into
spatial directions, orthogonal to the direction of ${\mathbf r}$.
Due to positivity, the potential is bound from below.\footnote{
The potential that is determined from Wilson loops depends on the
lattice cut-off, $a$, and can be factorised into a finite
potential $\hat{V}(r)$ and a (positive) self energy contribution:
$V(r;a)=\hat{V}(r)+V_{\mbox{\scriptsize self}}(a)$.
The latter diverges in the continuum limit 
(see Sec.~\ref{sec:pert}),
%\ref{sec:pertr}),
whereas the potential $\hat{V}(r)$ will become negative 
at small distances. Thus $V(r;a)$ is indeed bounded from below
by $V(0)$=0.}
Therefore,
convexity implies that $V(r)$ is a monotonically rising function of $r$,
{\it i.e.}
\begin{equation}
V'(r)\geq 0.
\end{equation}

In Ref.~\cite{Simon:1982yv}, which in fact preceded Ref.~\cite{Bachas:1986xs},
somewhat more strict upper and lower limits on Wilson loops,
calculated on a lattice, have been derived:
Let $a_{\sigma}$ and $a_{\tau}$
be temporal and spatial lattice resolutions.
The main result for rectangular Wilson loops
in representation $D$ and $d$ space-time dimensions then
is
\begin{equation}
\langle W(a_{\sigma},a_{\tau})\rangle^{rt/(a_{\sigma}a_{\tau})}\leq
\langle W(r,t)\rangle
\leq(1-c)^{r/a_{\sigma}+t/a_{\tau}-2},
\end{equation}
with
$c=\exp[-4(d-1)D\beta]$.
The resulting bounds
on $V(r)$ for $r>a_{\sigma}$ read
\begin{equation}
\label{eq:lowbound}
-\ln(1-c)\leq a_{\tau}V(r)\leq -\frac{r}{a_{\sigma}}\ln\langle W(a_{\sigma},a_{\tau})\rangle;
\end{equation}
in consistency with Ref.~\cite{Seiler:1978ur},
the potential (measured in lattice units $a_{\tau}$)
is bound from above by a linear function of $r$ and
it takes positive values everywhere.

\subsection{Strong coupling expansions}
Expectation
values can be approximated by expanding
the exponential of the lattice
action, in terms of the inverse coupling
$\beta=2N_C/g^2$, giving
$\exp(-\beta S)=1-\beta S+\cdots$.
This strong coupling expansion is similar to a
high temperature expansion in statistical mechanics. When the Wilson
action is used each factor $\beta$ is accompanied by a plaquette and
certain diagrammatic rules can be
derived~\cite{Wilson:1974skgb,Wilson:1976zj,Balian:1975xw,Creutz:1978yy,Drouffe:1983fv}.
Let us consider a strong coupling expansion of the Wilson loop,
Eq.~\ref{eq:wl}.
Since the integral over a single group element vanishes,
\begin{equation}
\int dU\,U=0,
\end{equation}
to zeroth
order, we have $\langle W\rangle = 0$. To the next order in $\beta$, it
becomes possible to cancel the link variables on the contour $\delta C$
of the Wilson
loop by tiling the whole minimal enclosed
(lattice) surface $C$
with plaquettes. Hence one
obtains the expectation
value~\cite{Creutz:1978yy,Creutz:1984mggb,Montvay:1994cygb,Smit:2002uggb}
\begin{equation}
\langle W(C)\rangle=\left\{\begin{array}{l}
\left[\beta/4\right]^{-\mbox{\scriptsize area}(\delta C)}+\cdots, \ \ N_C=2\\
\left[\beta/2N_{C}^2\right]^{-\mbox{\scriptsize area}(\delta C)}+\cdots, \ \ N_C>2
\end{array}\right\}
\end{equation}
for $SU(N_C)$ gauge theory. If we now consider the case of a rectangular
Wilson loop that extends $r/a$ lattice points into a spatial
and $t/a$ points into the temporal direction,
we find the area law,
\begin{equation}
\label{eq:areal}
\langle W({\mathbf r},t)\rangle = \exp\left[-\sigma_d rt\right]+
\cdots,
\end{equation}
with a string tension
\begin{equation}
\label{eq:str}
\sigma_da^2 = -d_r\,\ln \frac{\beta}{18}.
\end{equation}
The numerical value of the denominator applies to
$SU(3)$ gauge theory;
the potential is linear with slope $\sigma_d$,
and so colour sources are confined at strong coupling.
Here $d_r=(|r_1|+|r_2|+|r_3|)/r\geq 1$ denotes the ratio between lattice and
continuum norms and deviates from $d_r=1$ for source separations
${\mathbf r}$ that are not parallel to a lattice axis. The string tension
of Eq.~\ref{eq:str} depends on $d_r$ and, therefore,
on the lattice direction; $O(3)$ rotational
symmetry is explicitly
broken down to the cubic subgroup $O_h$. The extent of violation
will eventually be reduced as one increases $\beta$ and considers
higher orders of the expansion.
Such high order strong coupling expansions have indeed been performed
for Wilson loops~\cite{Munster:1980vk} and glueball
masses~\cite{Munster:1981es}.
Unlike standard perturbation theory,
whose convergence is known to be at best
asymptotic~\cite{Dyson:1952tj,Zinn-Justin:1981uk}, the strong coupling
expansion is analytic around
$\beta=0$~\cite{Osterwalder:1978pc} and, therefore, has
a finite radius of convergence.

\begin{figure}[b]
\centerline{\includegraphics[height=84mm]{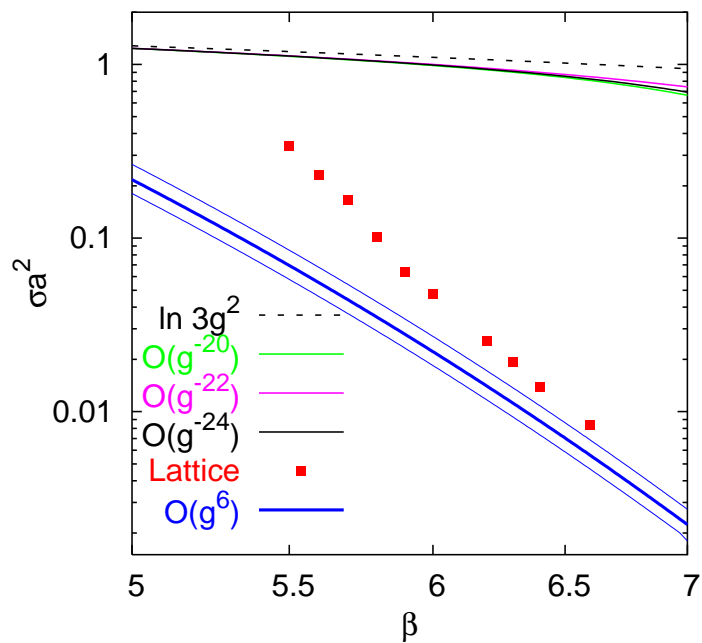}}
\vspace*{8pt}
\caption{Comparison of strong and weak coupling expansions with
non-perturbative results for the string tension.
The upper lines correspond to the strong coupling limit beginning with
$\ln 3g^2$ (dashed line) --- the lowest order term 
from Eq.~\protect\ref{eq:str} for $SU(3)$. The other upper lines
show the additional approximations up to $O(g^{-24})$. The lower lines
correspond to the weak coupling limit up to $O(g^{6})$}
\label{fig:crossover}
\end{figure}

Strong coupling $SU(3)$ gauge theory results 
seem to converge for \mbox{$\beta < 5$} --- see, 
for example, Ref.~\cite{Creutz:1984mggb}. One would have hoped
to eventually identify a crossover region of finite extent
between the validity regions of the strong and weak coupling
expansions~\cite{Kogut:1980pm}, or at least a transition point between
the leading order strong coupling behaviour, $a^2\propto-\ln(\beta/18)$,
of Eq.~\ref{eq:str}
and the weak coupling limit, $a^2\propto\exp[-2\pi\beta/(3\beta_0)]$,
set by the asymptotic freedom of the QCD $\beta$-function.
However, even after re-summing the strong coupling series
in terms of improved expansion parameters and applying sophisticated
Pad\'e approximation techniques~\cite{Smit:1982fx}, nowadays
such a \mbox{direct} crossover region does not appear to exist, necessitating
one to employ Monte Carlo simulation techniques.
In Fig.~\ref{fig:crossover} we compare the strong coupling expansion
for the string tension, calculated to 
$O(g^{-24})$ {\it i.e.} $O(\beta^{12})$~\cite{Munster:1980vk}
with results from lattice simulations. 
While at large
$\beta$ the lattice results approach the weak coupling limit, there appears
to be no overlap between weak and strong coupling and neither between
strong coupling and lattice results in the region of interest ($\beta\approx
6$ corresponds to a lattice spacing $a^{-1}\approx$ 2~GeV). We have
taken the $n_F=0$ value of the QCD $\Lambda$ parameter as determined
non-perturbatively in Ref.~\cite{Capitani:1998mq} as normalization for the weak
coupling expansion. The error band of the $O(g^6)$ expectation
is due to the corresponding
statistical uncertainty. The $O(g^4)$ central value lies within this band.
There is no normalization ambiguity in the strong
coupling results.
Also at $\beta>5$, the quality of convergence of the strong coupling expansion
diminishes.
This break-down might be related to a roughening
transition that is
discussed, for example, in Refs.~\cite{Luscher:1981ac,Drouffe:1981dp}.

We would like to remark that the area law of Eq.~\ref{eq:areal} is a rather
general result for strong coupling expansions in the fundamental representation
of compact gauge groups. In particular, it  applies also to $U(1)$ gauge theory
which we do not expect to confine in the continuum. In fact,
based on
duality arguments, Banks, Myerson and Kogut~\cite{Banks:1977cc} have
succeeded in proving
the existence of a confining phase in the four-dimensional theory and
suggested the existence of a phase transition while
Guth~\cite{Guth:1980gz} has proven that, at least in the
non-compact formulation
of $U(1)$, a Coulomb phase exists.
Indeed, in numerical simulations of (compact) $U(1)$ lattice gauge theory two
such distinct phases were found~\cite{Creutz:1979zg,Lautrup:1980xr},
a Coulomb phase
at weak coupling and a confining phase at strong coupling.
The question whether the confinement one finds in $SU(N_C)$
gauge theories in the strong coupling limit
survives the continuum limit, $\beta\rightarrow\infty$,
can at present only be answered by means of
numerical simulation (and has been answered positively).

\subsection{String picture}
\label{ch1.subs.string}
The infra-red properties of QCD might be reproduced by effective theories of
interacting strings. String models share many aspects with
the strong coupling expansion.
Originally, the string picture of confinement had
been discussed by Kogut and Susskind~\cite{Kogut:1975aggb}
as the strong coupling limit
of the Hamiltonian formulation of lattice QCD. The strong coupling
expansion of a Wilson loop can be cast into a sum of weighted random
deformations of the minimal area world sheet. This sum can
then be
interpreted to represent
a vibrating string. The physical picture behind such
an effective
string description is that of the electric flux between two colour
sources being squeezed into
a thin, effectively one-dimensional, flux tube or Abrikosov-Nielsen-Olesen
(ANO)
vortex~\cite{Abrikosov:1957sx,Nielsen:1973ve,'tHooft:1974jz,Migdal:1984gj}.
As a consequence, this yields a constant energy density
per unit length and a static
potential that is linearly rising as a function of the distance.

One can study the spectrum of such a vibrating string in simple
models~\cite{Luscher:1981ac,Luscher:1980fr,Luscher:1981iy}. Of course,
the string action is not {\em a priori} known. The simplest possible
assumption, employed in the above references,
is that the string is described by the Nambu-Goto
action~\cite{Goto:1971ce,Nambu:1974zg}
in terms of $d-2$ free bosonic fields
associated to the transverse degrees of freedom of the string.
In this picture, the static potential
is (up to a constant term) given by~\cite{Luscher:1980fr,Arvis:1983fp}
\begin{equation}
\label{eq:stringe}
V(r)=\sigma r\sqrt{1-\frac{(d-2)\pi}{12\,\sigma\,r^2}}=
\sigma\, r-\frac{(d-2)\pi}{24\,r}-\frac{(d-2)^2\pi^2}{1152\,\sigma\,r^3}
-\cdots.
\end{equation}
For a fermionic string~\cite{Caselle:1987ek}
one would expect
the coefficient of the correction term to the linear behaviour
to be only one quarter as big as the Nambu-Goto one above.
In the bosonic string picture,
excited levels are separated from the ground state by
\begin{equation}
\label{eq:excit}
V_n^2(r)=V^2(r)+(d-2)\pi n\sigma=\left[V(r)+\frac{(d-2)\pi n}{2r}-
\cdots\right]^2,
\end{equation}
with $n$ assuming integer values. 
It is clear from Eq.~\ref{eq:stringe} that
the string picture at best applies to distances
\begin{equation}
r\gg r_c=\sqrt{\frac{(d-2)\pi}{12\,\sigma}}.
\label{ch3.eq.rc}
\end{equation}
For $d=4$  one obtains
$r_c\approx 0.33$~fm, when using the value $\sqrt{\sigma}\approx 430$~MeV from
the $\rho,a_2,\cdots$ Regge trajectory.

The expectation of Eq.~\ref{eq:stringe} has been very accurately
reproduced in numerical simulations of $Z_2$ gauge theory in $d=3$ space-time
dimensions \cite{Caselle:1997ii,Caselle:2002ah}.
In contrast, for $d=4$ $SU(3)$ gauge theory the spectrum of
hybrid potentials still differs significantly from the
expectation of Eq.~\ref{eq:excit}
 for distances as large as 2~fm
~\cite{Morningstar:1998da,Juge:2004xr}.
However, qualitatively the string picture is supported by the $SU(3)$ data
too, since the hybrid potentials at large $r$ are found to group themselves
into various bands that are separated by approximately equidistant gaps.
These will eventually converge to
values $\pi/r$, at even larger distances than accessible at present.
Such an observation would support the
existence of a bosonic string description of confining gauge theories in the
very low energy
regime~\cite{Akhmedov:1996mw,Polyakov:1997nc,Chernodub:1998ie,Antonov:1998wi,Baker:1999xn}.
Of course, in $d<26$, the string Lagrangian is not
renormalisable --- but only effective --- and higher order correction terms 
like torsion and rigidity will in
general have to be added~\cite{Polchinski:1991ax}.

It is hard to disentangle in $d=4$
the (large distance) $1/r$ term, expected from string vibrations, from the
perturbative Coulomb term at short distances. However, a high precision
attempt has been made recently, with promising results \cite{Luscher:2002qv}.
As an alternative, three-dimensional
investigations (where perturbation theory yields a logarithmic contribution)
have been suggested~\cite{Ambjorn:1984yu,Majumdar:2002mr}.
Another way out is to determine
the mass of a
closed string, encircling a boundary of the lattice with a spatial
extent $l=L_{\sigma}a$ \mbox{(a torelon),}
which is not polluted by a perturbative tail.
The bosonic string expectation in this case would be~\cite{Ambjorn:1984yu}
\begin{equation}
\label{eq:tor}
E_n(l)=
\sigma\, l-\frac{(d-2)\pi}{6\,l}+\cdots.
\end{equation}
The na\"{\i}ve range of validity of the picture is
$l\gg l_c=2\,r_c\approx 0.66$~fm. The numerical value applies to
$d=4$ from Eq.~\ref{ch3.eq.rc}. An investigation of the finite size dependence of the
torelon mass in $d=4$ $SU(2)$ gauge theory has been performed some time ago
by
Michael and Stephenson \cite{Michael:1994ej} who found excellent agreement
--- on the 3~\% level --- with the bosonic string picture
already for distances $1\,\mbox{fm}\leq l\leq 2.4$~fm, quite close
to $l_c$. Qualitative agreement has also been reported by
Teper~\cite{Teper:1998te} from simulations of
$SU(2)$, $SU(3)$, $SU(4)$ and $SU(5)$ gauge theories in three dimensions
as well as in a recent study of four-dimensional $SU(2N)$ gauge theories
by Lucini {\em et al.}~\cite{Lucini:2004my}.

The bosonic string picture prediction of the free energy, calculated
from Polyakov line correlators,
at finite temperatures $T$ is
similar to Eq.~\ref{eq:tor}
\begin{equation}
-\frac{1}{\beta}\ln\langle P^*(r)P(0)\rangle
=\sigma(\beta)r+\cdots,\quad\sigma(T)=
\sigma-\frac{(d-2)\pi}{6}T^2+\cdots,
\end{equation}
with validity for $r\gg T=aL_{\tau}$ \cite{deForcrand:1985cz}.
The Polyakov line is defined as
\begin{equation}
\label{eq:pl2}
P({\mathbf x})=\Tr\,\left\{
{\mathcal T}\left[\exp\left(
i\int_0^{aL_{\tau}}\!dx_4\,A_4(x)\right)\right]\right\}=
\Tr\,\left(\prod_{x_4=0}^{aL_{\tau}}U_{x,4}\right),
\end{equation}
where ${\mathcal T}$ denotes time ordering of the argument.
The dependence of the effective string tension on the
temperature has been checked
for rather low $T^{-1}<1.24\,T_c^{-1}\approx 0.93\,\mbox{fm}$
in studies of $SU(3)$ gauge theory \cite{Kaczmarek:1999mm,Kaczmarek:2003ph}.
Although the sign of the leading correction term
to the zero temperature limit
is correct, the difference comes out to be larger than predicted.
It would be interesting to check whether the result will converge towards
the string expectation at lower temperatures.

\subsection{The potential in perturbation theory}
\label{sec:pert}
The strong coupling expansion is specific to the lattice
regularisation.\footnote{However, new strong coupling methods
have been developed in the large $N_C$ limit, based on the
Maldacena conjecture of QFT/AdS correspondence.}
However, the expectation value of a Wilson loop can also be
approximated  using standard perturbative techniques.

We will discuss the leading order weak coupling result that
corresponds to single gluon exchange between two static
colour sources which, although we neglect the spin structure,
we will call ``quarks'' for convenience. From the Lagrangian,
${\mathcal L}_{YM}=\frac{1}{2g^2}\gbTr F_{\mu\nu}F_{\mu\nu}$, one can easily
derive the propagator of a gluon with four-momentum $q$,
\begin{equation}
\label{eq:gluonprop}
G^{ab}_{\mu\nu}(q)=g^2\frac{\delta^{ab}\delta_{\mu\nu}}{q^2},
\end{equation}
where $\mu,\nu$ are Lorentz indices and $a,b=1,\cdots N_A$ label the
colour generators with $N_A^2=N_C^2-1$ for $SU(N_C)$. 
The same calculation can be done starting from a
lattice discretised action. The Wilson action
yields the result of Eq.~\ref{eq:gluonprop} up to the
replacement
\begin{equation}
q_{\mu}\rightarrow \hat{q}_{\mu}=\frac{2}{a}\sin
\left(\frac{aq_{\mu}}{2}\right).
\label{ch1.eq.qmuh}
\end{equation}
Other lattice actions yield slightly different results but
they all approach Eq.~\ref{eq:gluonprop} in the continuum limit,
$a\rightarrow 0$.
Momentum space potentials can be obtained from the
on-shell static quark anti-quark  scattering amplitude:
the gluon interacts with two static external currents pointing into
the positive and negative time directions,
$A^a_{\mu,\alpha\beta}=\delta_{\mu,4}T^a_{\alpha\beta}$
and $A_{\nu,\gamma\delta}^{\prime b}=-\delta_{\nu,4}T^b_{\gamma\delta}$.
Hence, we obtain the tree level interaction kernel
\begin{equation}
\label{eq:kern}
K_{\alpha\beta\gamma\delta}(q)=-\frac{g^2}{q^2}
T^a_{\alpha\beta}T^a_{\gamma\delta}.
\end{equation}

For sources in the fundamental representation,
the Greek indices denote the colour indices of the external currents
running from $1$ to $N_C$ and the quark anti-quark state can
be decomposed into two irreducible representations of $SU(N_C)$,
\begin{equation}
{\mathbf N_C}\otimes{\mathbf N_C^*}={\mathbf 1}\oplus{\mathbf N_A}.
\end{equation}
We can now either start from a singlet or
an octet\footnote{We call the state ${\mathbf N_A}$ an ``octet'' state,
having the group $SU(3)$ in mind.} initial
$\Phi_{\beta\gamma}=Q_{\beta}Q^*_{\gamma}$ 
state,
\begin{eqnarray}
\Phi^{\mathbf 1}_{\beta\gamma}&=&\delta_{\beta\gamma},\\
\Phi^{\mathbf N_A}_{\beta\gamma}&=&
\Phi_{\beta\gamma}-\frac{1}{N_C}\delta_{\beta\gamma},
\end{eqnarray}
where the normalisation is such that
$\Phi^i_{\alpha\beta}\Phi^{j}_{\beta\alpha}=\delta^{ij}$.
A contraction with the group generators of Eq.~\ref{eq:kern} yields
\begin{eqnarray}
\Phi^{\mathbf 1}_{\beta\gamma}T_{\alpha\beta}^aT_{\gamma\delta}^a
&=&C_F\Phi^{\mathbf 1}_{\alpha\delta},\\
\Phi^{\mathbf N_A}_{\beta\gamma}T_{\alpha\beta}^aT_{\gamma\delta}^a
&=&-\frac{1}{2N_C}\Phi^{\mathbf N_A}_{\alpha\delta},
\end{eqnarray}
where
$C_F=N_A/(2N_C)$
is the quadratic Casimir charge of the fundamental representation.

We end up with the potentials in momentum space,
\begin{equation}
V_s(q)=-C_Fg^2\frac{1}{q^2},\quad V_o(q)=\frac{g^2}{2N_C}\frac{1}{q^2}=
-\frac{1}{N_A}V_s(q),
\label{ch3.eq:potmom}
\end{equation}
governing interactions between fundamental charges coupled to a
singlet and to an octet,
respectively: the force in the singlet channel is attractive while that
in the octet channel is repulsive and smaller in size.

How are these potentials related to the static 
position space inter-quark potential,
defined non-perturbatively through the Wilson loop,
\begin{equation}
\label{eq:vdef}
V({\mathbf r})=-\lim_{t\rightarrow\infty}\frac{d}{dt}
\ln\langle W({\mathbf r},t)\rangle?
\end{equation}
The quark anti-quark state creation operator, $\Psi_{\mathbf r}$,
within the Wilson loop 
contains a gauge transporter and couples to the gluonic degrees of freedom.
Thus, in general,
it will have overlap with both, $Q Q^*$ singlet and octet
channels\footnote{Of
course, for a quark and anti-quark being at different 
spatial positions, the singlet-octet classification
should be consumed with caution in a non-perturbative
context.}.
Since the singlet channel is energetically preferred, {\it i.e.} $V_s<V_o$,
we might expect the static potential to correspond to the singlet
potential. Up to order $g^6$ this is indeed the case:
to lowest order, the Wilson loop --- defined by the closed contour 
$\delta C$ --- is given by
the Gaussian integral
\begin{equation}
\label{eq:wile}
\langle W({\mathbf r},t)\rangle=
\exp\left\{-\frac{1}{2}\int\! d^4x\,d^4y
J^a_{\mu}(x)
G^{ab}_{\mu\nu}(x-y)J^b_{\nu}(y)\right\},
\end{equation}
where $J_{\mu}^a=\pm T^a$ if $(x,\mu)\in\delta C$ and $J_{\mu}^a=0$
elsewhere\footnote{Note that this formula, which automatically
accounts for multi-photon exchanges, is exact in non-compact QED
(excluding fermion loops) to any order of
perturbation theory. However, in theories containing more complicated vertices,
like non-Abelian gauge theories or
compact lattice $U(1)$ gauge theory, correction terms have to be
added at higher orders in $g^2$.}.
Eq.~\ref{eq:wile} implies for $t\gg r$
\begin{equation}
\label{eq:wilint}
\langle W({\mathbf r},t)\rangle=
\exp\left(C_Fg^2t\int_{-t/2}^{t/2}\!dt'\ [G({\mathbf r},t')-G({\mathbf 0},t')]
\right).
\end{equation}
We have omitted gluon exchanges between the spatial closures
of the Wilson loop from the above formula. Up to order $g^6$
(two loops), such contributions result in terms
whose exponents are proportional to $r$ and $r/t$
and, therefore, do not affect the potential
of Eq.~\ref{eq:vdef}.
The propagator $G^{ab}_{\mu\nu}(x)$, the Fourier transform of
$G^{ab}_{\mu\nu}(q)$ in Eq.~\ref{eq:gluonprop}, contains the function
\begin{equation}
G(x)=\int\frac{d^4q}{(2\pi)^4}\frac{e^{iqx}}{q^2},
\quad\int_{-\infty}^{\infty}\!dx_4\,G(x)=\frac{1}{4\pi}\frac{1}{r}.
\end{equation}
After performing the $t$-integration, we obtain
\begin{equation}
\label{eq:vfundpert}
V({\mathbf r},\mu)=-C_F\frac{\alpha_s}{r} + V_{\mbox{\scriptsize self}}(\mu),
\end{equation}
where $\alpha_s=g^2/(4\pi)$. The piece
\begin{equation}
\label{eq:self1}
V_{\mbox{\scriptsize self}}(\mu)=C_Fg^2\int_{q\leq\mu}
\frac{d^3q}{(2\pi)^3}\frac{1}{q^2}=C_F\alpha_s\frac{2}{\pi}\mu,
\end{equation}
that linearly diverges with the ultra-violet cut-off $\mu$,
results from self-inter\-act\-ions of the static (infinitely heavy)
sources.
Comparing Eqs.~\ref{ch3.eq:potmom} and \ref{eq:vfundpert}
we indeed find
\begin{equation}
\label{eq:equal}
V(q)=V_s(q),
\end{equation}
where
\begin{equation}
V({\mathbf q},0)=
\int\!d^3r\,e^{i{\mathbf q}\cdot{\mathbf r}} \hat{V}({\mathbf r}),\quad
\hat{V}({\mathbf r}) = V({\mathbf r},\mu)
-V_{\mbox{\scriptsize self}}(\mu).
\end{equation}
This self-energy ``problem'' is well known on the lattice
and has also received attention in continuum QCD, in the context of
renormalon ambiguities in quark mass
definitions~\cite{Bigi:1994em,Beneke:1998rk}.

Note that while $V_s$ corresponds to the static potential, the perturbation
theory relevant for hybrid excitations of the ground state potential
corresponds to $V_o$~\cite{Bali:2003jq}.

At order $\alpha_s^4$ a class of diagrams appears in a perturbative
calculation of the Wilson loop that results in
contributions to the static potential that diverge logarithmically
with the interaction time~\cite{Appelquist:1977tw}.
In Ref.~\cite{Brambilla:1999qa}, within the framework of effective
field theories, this effect has been related to ultra-soft
gluons due to which an extra scale, $V_o-V_s$,
is generated. Moreover, a systematic procedure has been
suggested to isolate and subtract such terms to obtain
 finite singlet and octet interaction potentials between heavy quarks.

The logarithmic divergence is related to the fact that
Eq.~\ref{eq:wilint} contains
an integration over the interaction time. For large times
and any fixed distance $r$, Wilson loops will decay exponentially
with $t$. However, the tree level propagator in position space
is proportional to $(r^2+t^2)^{-1}$, \mbox{{\it i.e.}  asymptotically}
decays like $t^{-2}$ only. We notice that
the integral receives significant contributions from the region of large $t$
as demonstrated by the finite $t\gg r$ tree level result
\begin{equation}
-\ln\langle W(r,t)\rangle=-\frac{C_F\alpha_s}{r}t\frac{2}{\pi}
\left\{
\arctan\frac{t}{r}-\frac{r^2}{2t}\left[
\ln\left(1+\frac{t^2}{r^2}\right)\right]\right\}
+(r+t)V_{\mbox{\scriptsize self}}.
\end{equation}

The tree level lattice potential can easily be obtained by replacing
$q_{\mu}$ by $\hat{q}_{\mu}$ from Eq.~\ref{ch1.eq.qmuh} 
and (in the case of finite lattice volumes)
the integrals by discrete sums over lattice momenta,
\begin{equation}
q_i=\frac{2\pi}{L_{\sigma}}\,\frac{n_i}{a},\quad
n_i=-\frac{L_{\sigma}}{2}+1,\cdots,\frac{L_{\sigma}}{2}.
\end{equation}
The lattice potential reads
\begin{equation}
\label{eq:latpot}
V({\mathbf r})=V_{\mbox{\scriptsize self}}(a)-C_F\alpha_s\left[\frac{1}{{\mathbf r}}\right],
\end{equation}
where
\begin{equation}
\label{eq:latpr}
\left[\frac{1}{{\mathbf r}}\right]=\frac{4\pi}{L_{\sigma}^3a^3}
\sum_{{\mathbf q}\neq {\mathbf 0}}\frac{e^{i{\mathbf qR}}}{\sum_i
\hat{q}_i\hat{q}_i}
\end{equation}
and $V_{\mbox{\scriptsize self}}(a)=C_F\alpha_s\left[1/{\mathbf 0}\right]$.
We have neglected the zero mode contribution that is suppressed
by the inverse volume $(aL_{\sigma})^{-3}$. In the continuum limit,
$[1/{\mathbf r}]$ approaches $1/r$ up to quadratic lattice artefacts
whose coefficients depend on the direction of ${\mathbf r}$ while
$V_{\mbox{\scriptsize self}}(a)$ diverges like
\begin{equation}
\label{eq:self2}
V_{\mbox{\scriptsize self}}(a)=C_F\alpha_s a^{-1}\times 3.1759115\cdots.
\end{equation}
The numerical value applies to the limit, $L_{\sigma}\rightarrow\infty$.
Note that under the substitution $\mu\approx 1.5879557\,\pi/a$,
Eq.~\ref{eq:self2} is identical to Eq.~\ref{eq:self1}.
One loop computations of on-axis lattice
Wilson loops can be
found in Refs.~\cite{Curci:1984wh,Wohlert:1985hk,Heller:1985hx},
while off-axis separations in QCD with and without sea quarks
have been realized in Ref.~\cite{Bali:2002wf}.
The tree level
form, Eq.~\ref{eq:latpot}, is often employed to parameterise
lattice artefacts --- see  Subsec.~5.9.3.1 of Chapter~5.
\section{Quark-Antiquark Potentials between Non-Static \mbox{Quarks}}
\label{sec:gluon}
\footnote{At this point the author ``ran-out-of-steam" and so, because
of the pressure of time, the editor felt that this chapter should be 
rapidly concluded --- a task carried out by the editor himself with
the semi-approval of the author.}
So far in this chapter the r\^{o}le of the interquark potential has been
to test various limits and models for QCD. For example, in 
Fig.~\ref{fig:crossover} a comparison was made between the weak and
strong coupling limits of QCD and also with the corresponding lattice
results. Likewise, in Subsec.~\ref{ch1.subs.string} string models were
compared with the infra-red properties of QCD. This  r\^{o}le is in
stark contrast to that of the Nucleon-Nucleon potential [$V(NN)$] in
nuclear physics, where $V(NN)$ is mainly used as a stepping stone to 
the understanding of multi-nucleon systems. To this end, the many
parameters needed to define $V(NN)$ are first adjusted, more or less
freely, to fit two nucleon experimental data, with --- in some cases ---
values being imposed from meson-nucleon data or theories. 
Unlike $V(Q\bar{Q})$, the NN-potential is not able to predict  reliable
quantitative information about the input parameters to its theory --- the only
exception perhaps being an estimate of the $\pi NN$ coupling 
constant~\cite{ch1.NN.Stoks}. The reason why the form of $V(NN)$ is so 
complicated compared with that so far discussed for $V(Q\bar{Q})$, 
{\it i.e.} essentially $V_{Q\bar{Q}}(r)=-e/r+cr$, is 
because the latter is the interaction between two {\it static} quarks.
If, however, we go away from this limit, then immediately the spins of
the quarks begin to play a r\^{o}le resulting in forms similar to those 
encountered in $V(NN)$. These potentials should be reliable for
describing $b\bar{b}, \ b\bar{c}$ and $c\bar{c}$ states, 
since the $b$ and $c$ quarks are
still sufficiently heavy ($\approx 4.5$ and 1.2 GeV respectively) to not
require a relativistic treatment. However, as soon as $s, \ d, \ u$
quarks are involved, relativistic effects become important and even the
whole concept of an interquark potential should be questioned.
Furthermore,, the hope that this potential between {\it two} quarks
can account for multiquark systems has yet to be justified. Even so, 
this has not deterred its use as an effective interaction --- a topic
discussed in Chapter~5.

\subsection{Radial form of $V(Q\bar{Q})$} 
What is the form of $V(Q\bar{Q})$? 
To this question there is no unique answer, since --- as with
meson-exchange models of $V(NN)$ --- forms depend on the theory
from which the potential is derived and in which it should be utilized.
A good example of this ambiguity is the momentum dependence of the
potential. Even though the correct relativistic scattering
equation for two quarks is  the  Bethe-Salpeter equation,
for practical reasons, this needs to be simplified to, say, the 
semi-relativistic Blankenbecler--Sugar or non-relativistic 
Schr\"{o}dinger equations --- see Subsec.~5.1.2.2 of Chapter~5 for a
very brief discussion of this. 
If the basic Blankenbecler-Sugar equation is used directly then
the appropriate potential contains relativistic factors of the form
$E_Q=\sqrt{M^2+p^2}$, where $p$ is the momentum of the quark with 
mass $M$.
However, it is often convenient to expand these momentum
factors in powers of $p/M$ resulting in a potential 
appropriate for a  Lippmann--Schwinger or  Schr\"{o}dinger approach.

\begin{figure}[ht] 
\centering 
\includegraphics[height=0.55\textwidth]{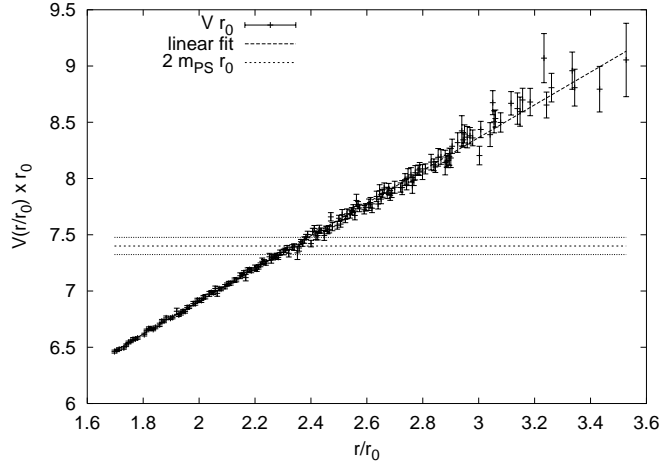} 
\caption{The linearly rising confinement as calculated on a 
lattice  \protect\cite{ch1:Bolder00}
compared with $V_C(Mm)=cr$ in Eq.~\protect\ref{eq:ch1.vc}.
\label{ch1:Bolderfig}} 
\end{figure}

Just as the interaction between two static quarks contained two distinct
parts --- {\it i.e.} the gluon exchange term $V_G=-\frac{4}{3}\alpha_s/r$
and the confining 
term $V_C=c r$ --- so can the interaction between a  heavy quark of
mass $M$ and an anti-quark of mass $m$ be expressed as 
\begin{equation}
\label{eq:ch1.vc+vg}
V(Mm)=V_G(Mm)+V_C(Mm),
\end{equation}
where
\[V_G(Mm)(-\frac{4}{3}\alpha_s)^{-1}=
\frac{1}{r}
-\frac{2\pi}{3Mm}\delta^{(3)}(r) \bm{ \sigma_M.\sigma_m}
-\frac{1}{4Mmr^3}S_{12}\]
\begin{equation}  
\label{eq:ch1.vg}
-\frac{1}{2r^3}\left[ \frac{M^2+m^2}{2M^2m^2}+\frac{2}{Mm}\right]
\bm{ L.S}
+\frac{1}{8r^3}\frac{M^2-m^2}{M^2m^2}
(\bm{ \sigma}_M-\bm{\sigma}_m).\bm{ L}+\cdots,
\end{equation}
obtained  by a non-relativistic reduction of the
one-gluon-exchange mechanism and 
\begin{equation}
\label{eq:ch1.vc}
V_C(Mm)=cr
-\frac{c}{r}\frac{M^2+m^2}{4M^2m^2}\bm{ L.S}
+\frac{c}{r}\frac{M^2-m^2}{8M^2m^2}(\bm{\sigma}_M-\bm{\sigma}_m).\bm{L}
+\cdots.
\end{equation}
These expressions for $V_G(Mm)$ and $V_C(Mm)$ do not include effects from
expanding the \mbox{$E=\sqrt{M^2+p^2}$} terms mentioned above
--- see, for example, Ref.~\cite{ch1.QQdelreg}. Also here, for
simplicity, they do not include
spin-independent terms proportional to $p^2$ and
$\delta^{(3)}(r)$, since their form becomes  ambiguous when going from a
momentum space to a coordinate space representation. It should be added
that it is convenient to have $V_G(Mm)$ in coordinate space, since it is
hard to deal with the linearly rising term $cr$ from $V_C(Mm)$ in
momentum space. 
In addition to this ambiguity there are three others that enter in practice:
\begin{enumerate}
\item A crucial term in Eq.\ref{eq:ch1.vg} is the hyperfine
interaction proportional to $\bm{\sigma_M . \sigma_m}$, which splits the
energies of pseudoscalar and vector mesons. However, it is seen that
it is proportional to $\delta(\bf{r})$ --- a radial form that needs to be
regulated before it can be used in a wave-equation. This can be
accomplished in a variety of ways , of which the most simple is to 
make the replacement
\begin{equation}
\label{eq:ch1.God.Is}
\delta({\bf r})\rightarrow \frac{a^3}{\pi^{3/2}}\exp(-a^2r^2)
\end{equation}
as was done in the original work of Godfrey and 
Isgur~\cite{ch1.QQpot.God.Is}. Another approach is to consider
this term as an effective interaction that is added in first order 
perturbation theory using those wavefunctions generated by the rest of
$V(Q\bar{Q})$. In this case the overall constant is considered to be
essentially a free parameter although reasonable estimates can be made
from models involving instantons~\cite{ch1.instpot}.

\item In Eq.~\ref{eq:vfundpert},  $\alpha_s$ is the quark-gluon running
coupling "constant" {\it i.e.} it depends on momentum $k$. This can be
 parameterised in a variety of ways {\it e.g.} from 
Ref.~\cite{ch1.QQreg}
\begin{equation}
\label{ch1:eq.alphapar}
\alpha_s(k^2)=\frac{12\pi}{27}\frac{1}{\ln[(k^2+4m_g^2)/\Lambda^2]},
\end{equation}
where there are two parameters 

i) the gluon effective mass parameter $m_g\approx 240$ MeV and

ii) the QCD scale parameter $\Lambda \approx 280$ MeV.

In Ref.~\cite{ch1.QQdelreg} this is combined with the regulator of 
$\delta({\bf r})$ by the replacement
\begin{equation}
\alpha_s\delta({\bf r})\rightarrow \frac{1}{2\pi^2r}\int_0^{\infty}
dk k \sin(kr)\left( \frac{M+m}{E_M+E_m}\right)
\left(\frac{Mm}{E_ME_m}\right)\alpha_s(k^2),
\end{equation}
where $E_M=\sqrt{M^2+(k^2/4)}$ and $E_m=\sqrt{m^2+(k^2/4)}$.
\item In the above it is implicitly assumed that $V_C(Mm)$ is
purely \mbox{{\it scalar}.} However, there are reasons --- both theoretical
and phenomenological --- suggesting that there could be a sizeable
{\it vector} \mbox{component.} For example, the spin-orbit splitting
from $V_G(Mm)$ is $\propto 1/r^3$, whereas that in $V_C(Mm)$ is
$\propto -1/r$. This indicates that the natural spin-orbit
ordering of a coulomb-like potential should be inverted for high
partial waves --- a feature not seen experimentally or in
lattice calculations. This is discussed  in Subsec.~5.9.3.2 in Chapter~5. 
\end{enumerate}
Lattice calculations are able to isolate several of the components in
the potential $V(Mm)$ in Eq.~\ref{eq:ch1.vc+vg}. For example, the leading
confinement term $V_C(Mm)=cr$ of Eq.~\ref{eq:ch1.vc} is shown  in 
Fig.~\ref{ch1:Bolderfig}. 
Another point of interest in this figure is that there is no sign of the
expected
flattening at $r/r_0\approx 2.4$, where it becomes energetically 
favourable to create two mesons --- denoted by the 
horizontal band \cite{ch1:Bolder00}. For other components of
the potential the lattice data is less precise, but still there is
reasonable agreement with the expectations from Eqs.~\ref{eq:ch1.vg} and 
\ref{eq:ch1.vc}. Probably the most extensive lattice study of the various
forms contained
in  the potential appear in Ref.~\cite{Bali:1997am}.

\subsection{Comparison with the form of $V(NN)$}
For those more familiar with internucleon potentials it may be of interest
to compare the forms that appear in $V(NN)$ with those  in $V(Q\bar{Q})$.
This is, probably, best done by writing down explicitly the $\omega$-meson
contribution to the familiar One-Boson-Exchange-Potential, since this
meson is the one that resembles most closely a gluon  both being vector
particles independent of isospin/flavour.
In $V(NN)$, the $\omega$-meson contributes much of the short range repulsion
and also has a strong Spin-orbit potential. 
The form given in Ref.~\cite{ch1.BrysNN} is
\[\frac{V_{\omega}}{g^2}=
\frac{\exp(-\bar{m} r)}{r}-\frac{1}{2M^2}\left[\bm{\nabla}^2
\frac{\exp(-\bar{m} r)}{ r}+\frac{\exp(-\bar{m}r)}{r}\bm{\nabla}^2\right]\]
\[+\frac{1}{2M^2}\left[\bar{m}^2\frac{\exp(-\bar{m}r)}{r}
-4\pi\delta^{(3)}(r)\right]
(1+\frac{\bm{ \sigma_1.\sigma_2}}{3})\]

\begin{equation}
\label{ch1:eq.BS}
-\frac{\bar{m}^2}{4M^2}\left[\frac{1}{3}+\frac{1}{\bar{m} r}+
\frac{1}{(\bar{m} r)^2}\right]
\frac{\exp(-\bar{m}r)}{ r}S_{12}
+\frac{3}{2}\frac{1}{M^2}\frac{1}{r}\frac{d}{dr}
\left[\frac{\exp(-\bar{m} r)}{ r}\right]\bm{L.S},
 \end{equation}
where $M$ is the nucleon mass and  $\bar{m}=783$ MeV $\approx 4$fm$^{-1}$
is the mass  of the
$\omega$. When $\bar{m}\rightarrow 0$ to compare with one-gluon-exchange,
 this reduces to
\[\frac{V(m\rightarrow 0)}{g^2}=\frac{1}{r}-\frac{1}{2M^2}
\left[\bm{\nabla}^2 \frac{1}{ r}+\frac{1}{r}\bm{\nabla}^2\right]
-\frac{2\pi}{M^2}\delta^{(3)}(r)\left[1+\frac{\bm{
\sigma_1.\sigma_2}}{3}
\right]\]
\begin{equation}
\label{ch1:eq.BS0}
-\frac{1}{4M^2}\frac{1}{r^3}S_{12}-
\frac{3}{2}\frac{1}{M^2}\frac{1}{r^3}\bm{L.S}.
\end{equation} 
This form is now quite similar to 
the one-gluon-exchange potential $V_G$ in Eq.~\ref{eq:ch1.vg} ---
the only differences being the appearance of the spin-independent terms 
containing $\nabla^2$ and $\delta^{(3)}(r)$, which are not unique in
going from momentum space to coordinate space. Of course, it is 
possible to stay throughout in momentum space, since --- unlike
$V_C(Mm)$ --- here there is no linearly rising potential $cr$. 
However,  there are also differences in how Eqs.~\ref{eq:ch1.vg} and
\ref{ch1:eq.BS} are treated. Normally in Eq.~\ref{ch1:eq.BS}
the $\delta^{(3)}(r)$ terms do not appear, since there are 
physically motivated models of form factors at the $NN\omega$-vertices,
which  smooth out such singularities. Also the results are not
crucially dependent on these form factors since the main r\^{o}le of the
$\omega$-meson is to generate a strong spin-isospin independent {\it repulsion}
that essentially excludes the $NN$-wavefunction from the region of the origin.
In contrast, the one-gluon-exchange potential  in Eq.~\ref{eq:ch1.vg}
has an {\it attractive} spin-isospin independent term 
$(-\frac{4}{3}\alpha_s/r)$. This means that  uncertainties in the
parametrizations in Eqs.~\ref{eq:ch1.God.Is} and \ref{ch1:eq.alphapar}
are more important.

\section{Conclusions}
This chapter has attempted to describe a few of the many topics that
could be covered by the title. It has not aimed at being in anyway
comprehensive --- the selection being rather subjective with the
following ideas in mind. 
%\begin{enumerate}\item 

\vspace{0.3cm}

\noindent 1) The chapter began and ended with the theme
that the interquark potential $V(Q\bar{Q})$ could be viewed as a bridge
for modelling QCD with somewhat the same r\^{o}le that the nucleon-nucleon
potential $V(NN)$ plays in nuclear physics. To further bring out the analogy
 between $V(Q\bar{Q})$ and $V(NN)$ the radial forms were compared and
contrasted in Sec.~\ref{sec:gluon}. However, it should be added that
this possible ``interdisciplinary bridge" is not well understood
and its study is rather neglected. 

\vspace{0.3cm}

\noindent 2) The potential $V(Q\bar{Q})$ can be used for testing approximations
to QCD and its lattice formulation. This is illustrated in
Fig.~\ref{fig:crossover}, where it is shown that the weak and strong
coupling limits  {\it do not} overlap for the energy range of 
interest  {\it i.e.} between 100 MeV and a few GeV. 
Furthermore, over this range neither limit agrees with the 
corresponding (non-perturbative) lattice calculation. This feature shows
that we are ``forced" into performing lattice calculations, since there
seems to be no other way of treating QCD with the couplings of most
interest.
%\end{enumerate}

\vspace{0.3cm}

Here no mention has been made about deriving a discretized form of $V(Q\bar{Q})$.
This topic has been the scene of much activity and comes under the 
heading of Non-Relativistic QCD (NRQCD). Here an expansion is made
in terms of $\Lambda/M_Q$ with $\Lambda\sim 1$ GeV  being a characteristic energy
scale  for non-perturbative effects. A few comments can be found in 
Subsec.~5.1.2.1 of Chapter~5.
\section*{Acknowledgements}
\addcontentsline{toc}{section}{Acknowledgements}
I express sincere gratitude to Tony Green, for his initiative
in suggesting this book and in particular for his patience in waiting
for a ridiculously long time for my contribution to arrive.
Discussions with Nora Brambilla, Antonio Pineda, Joan Soto and
Antonio Vairo are acknowledged.
I received support from a PPARC Advanced
Fellowship (grant PPA/A/S/2000/00271) as well as by PPARC grant
PPA/G/0/2002/0463.
%\bibliography{biblio}

\begin{thebibliography}{00}
%\section*{Bibliography}
\markright{Bibliography} 

\bibitem{Wegner:1984qt}
F.~J. Wegner,
\newblock {\it J. Math. Phys.}, 12:2259, 1971.

\bibitem{Wilson:1974skgb}
K.~G. Wilson,
\newblock {\it Phys. Rev.}, D10:2445, 1974.

\bibitem{Brown:1979ya}
L.~S. Brown and W.~I. Weisberger,
\newblock {\it Phys. Rev.}, D20:3239, 1979.

\bibitem{Eichten:1979pu}
E.~Eichten and F.~L. Feinberg,
\newblock {\it Phys. Rev. Lett.}, 43:1205, 1979.

%\bibitem{Eichten:1981mw}
E.~Eichten and F.~L. Feinberg,
\newblock {\it Phys. Rev.}, D23:2724, 1981.

\bibitem{Barc12} A. Barchielli, E. Montaldi and G.~M. Prosperi,
\newblock {\it Nucl.Phys.}, B296:625,1988, 
\newblock {\it erratum}, {\it ibid.}  B303:752,1988 

A. Barchielli, N. Brambilla and G.~M.  Prosperi,
\newblock {\it Nuovo Cim.} A103:59,1990 
\bibitem{Pineda:2000sz}
A. Pineda and A. Vairo
\newblock {\it Phys. Rev.}, D63:054007, 2001.

\bibitem{Chen:1995dg}
Yu-Qi Chen, Yu-Ping Kuang, and R.~J. Oakes,
\newblock {\it Phys. Rev.}, D52:264, 1995.

\bibitem{Bali:1997am}
G.~S. Bali, K. Schilling, and A. Wachter,
\newblock {\it Phys. Rev.}, D56:2566, 1997.

\bibitem{Brambilla:2000gk}
N. Brambilla, A. Pineda, J. Soto, and A. Vairo,
\newblock {\it Phys. Rev.}, D63:014023, 2001


\bibitem{Landau:1987qm}
L.~D. Landau and E.~M. Lifschitz.
\newblock {\it Lehrbuch der theoretischen Physik, Band 3: Quantenmechanik},
\newblock Akademie Verlag, Berlin, DDR, 1979.

\bibitem{Seiler:1978ur}
E.~Seiler,
\newblock {\it Phys. Rev.}, D18:482, 1978.

\bibitem{Osterwalder:1973dx}
K. Osterwalder and R. Schrader,
\newblock {\it Commun. Math. Phys.}, 31:83, 1973.

\bibitem{Osterwalder:1975tc}
K. Osterwalder and R. Schrader,
\newblock {\it Commun. Math. Phys.}, 42:281, 1975.

\bibitem{Bachas:1986xs}
C. Bachas,
\newblock {\it Phys. Rev.}, D33:2723, 1986.

\bibitem{Simon:1982yv}
B.~Simon and L.~G. Yaffe,
\newblock {\it Phys. Lett.}, 115B:145, 1982.

\bibitem{Wilson:1976zj}
K.~G. Wilson,
\newblock {\it Phys. Rept.}, 23:331, 1976.

\bibitem{Balian:1975xw}
R.~Balian, J.~M. Drouffe, and C.~Itzykson,
\newblock {\it Phys. Rev.}, D11:2104, 1975,
\newblock {\it  erratum}, {\it  ibid.}  D19:2514, 1979.

\bibitem{Creutz:1978yy}
M. Creutz,
\newblock {\it Rev. Mod. Phys.}, 50:561, 1978.

\bibitem{Drouffe:1983fv}
J. M. Drouffe and J. B. Zuber,
\newblock {\it Phys. Rept.}, 102:1, 1983.

\bibitem{Creutz:1984mggb}
M.~Creutz,
\newblock {\it Quarks, Gluons and Lattices}.
\newblock Cambridge University Press, Cambridge, UK, 1983.

\bibitem{Montvay:1994cygb}
I.~Montvay and G.~M{\"u}nster,
\newblock {\it Quantum fields on a lattice}.
\newblock Cambridge University Press, Cambridge, UK, 1994.

\bibitem{Smit:2002uggb}
J.~Smit,
\newblock Introduction to quantum fields on a lattice: A robust mate,
\newblock {\it Cambridge Lect. Notes Phys.}, 15:1, 2002.

\bibitem{Munster:1980vk}
G. M{\"u}nster and P. Weisz,
\newblock {\it Phys. Lett.}, 96B:119, 1980.
\newblock erratum, {\it  ibid.}  100B:519, 1981.

\bibitem{Munster:1981es}
G. M{\"u}nster,
\newblock {\it Nucl. Phys.}, B190:439, 1981.
\newblock errata, {\it  ibid.}  B200:536, 1982 and  B205:648, 1982.

\bibitem{Dyson:1952tj}
F.~J. Dyson,
\newblock {\it Phys. Rev.}, 85:631, 1952.

\bibitem{Zinn-Justin:1981uk}
J.~Zinn-Justin,
\newblock {\it Phys. Rept.}, 70:109, 1981.

\bibitem{Osterwalder:1978pc}
K.~Osterwalder and E.~Seiler,
\newblock {\it Ann. Phys.}, 110:440, 1978.

\bibitem{Kogut:1980pm}
J.~B. Kogut and J. Shigemitsu,
\newblock {\it Phys. Rev. Lett.}, 45:410, 1980.

\bibitem{Smit:1982fx}
J. Smit,
\newblock {\it Nucl. Phys.}, B206:309, 1982.

\bibitem{Capitani:1998mq}
S. Capitani, M. L{\"u}scher, R. Sommer, and H. Wittig,
\newblock {\it Nucl. Phys.}, B544:669, 1999.

\bibitem{Luscher:1981ac}
M.~L{\"u}scher,
\newblock {\it Nucl. Phys.}, B180:317, 1981.

\bibitem{Drouffe:1981dp}
J.~M. Drouffe and J.~B. Zuber,
\newblock {\it Nucl. Phys.}, B180:264, 1981.

\bibitem{Banks:1977cc}
T.~Banks, R.~Myerson, and J.~Kogut,
\newblock {\it Nucl. Phys.}, B129:493, 1977.

\bibitem{Guth:1980gz}
A.~H. Guth,
\newblock {\it Phys. Rev.}, D21:2291, 1980.

\bibitem{Creutz:1979zg}
M. Creutz, L. Jacobs, and C. Rebbi,
\newblock {\it Phys. Rev.}, D20:1915, 1979.

\bibitem{Lautrup:1980xr}
B.~Lautrup and M.~Nauenberg,
\newblock {\it Phys. Lett.}, 95B:63, 1980.

\bibitem{Kogut:1975aggb}
J. Kogut and L. Susskind,
\newblock {\it Phys. Rev.}, D11:395, 1975.

\bibitem{Abrikosov:1957sx}
A.~A. Abrikosov,
\newblock {\it Sov. Phys. JETP}, 5:1174, 1957.

\bibitem{Nielsen:1973ve}
H.~B. Nielsen and P.~Olesen,
\newblock {\it Nucl. Phys.}, B61:45, 1973.

\bibitem{'tHooft:1974jz}
G.~'t~Hooft,
\newblock {\it Nucl. Phys.}, B72:461, 1974.

\bibitem{Migdal:1984gj}
A.~A. Migdal,
\newblock {\it Phys. Rept.}, 102:199, 1983.

\bibitem{Luscher:1980fr}
M.~L{\"u}scher, K.~Symanzik, and P.~Weisz,
\newblock {\it Nucl. Phys.}, B173:365, 1980.

\bibitem{Luscher:1981iy}
M.~L{\"u}scher, G.~M{\"u}nster, and P.~Weisz,
\newblock {\it Nucl. Phys.}, B180:1, 1981.

\bibitem{Goto:1971ce}
T. Goto,
\newblock {\it Prog. Theor. Phys.}, 46:1560, 1971.

\bibitem{Nambu:1974zg}
Y.~Nambu,
\newblock {\it Phys. Rev.}, D10:4262, 1974.

\bibitem{Arvis:1983fp}
J.~F. Arvis,
\newblock {\it Phys. Lett.}, 127B:106, 1983.

\bibitem{Caselle:1987ek}
M.~Caselle, R.~Fiore, and F.~Gliozzi,
\newblock {\it Phys. Lett.}, B200:525, 1988.

\bibitem{Caselle:1997ii}
M.~Caselle, R.~Fiore, F.~Gliozzi, M.~Hasenbusch, and P.~Provero,
\newblock {\it Nucl. Phys.}, B486:245, 1997.

\bibitem{Caselle:2002ah}
M.~Caselle, M.~Hasenbusch, and M.~Panero,
\newblock {\it JHEP}, 01:057, 2003.

\bibitem{Morningstar:1998da}
C.~J. Morningstar, K.~J. Juge, and J.~Kuti,
\newblock {\it Nucl. Phys. Proc. Suppl.}, 73:590, 1999.

\bibitem{Juge:2004xr}
K.~Jimmy Juge, J.~Kuti, and C.~Morningstar,
\newblock QCD string formation and the Casimir energy,
\newblock hep-lat/0401032 

\bibitem{Akhmedov:1996mw}
E.~T. Akhmedov, M.~N. Chernodub, M.~I. Polikarpov, and M.~A. Zubkov,
\newblock {\it Phys. Rev.}, D53:2087, 1996.

\bibitem{Polyakov:1997nc}
A.~M. Polyakov,
\newblock {\it Nucl. Phys.}, B486:23, 1997.

\bibitem{Chernodub:1998ie}
M.~N. Chernodub and D.~A. Komarov,
\newblock {\it JETP Lett.}, 68:117, 1998.

\bibitem{Antonov:1998wi}
D. Antonov and D. Ebert,
\newblock {\it Phys. Lett.}, B444:208, 1998.

\bibitem{Baker:1999xn}
M.~Baker and R.~Steinke,
\newblock {\it Phys. Lett.}, B474:67, 2000.

\bibitem{Polchinski:1991ax}
J. Polchinski and A. Strominger,
\newblock {\it Phys. Rev. Lett.}, 67:1681, 1991.

\bibitem{Luscher:2002qv}
M. Luscher and P. Weisz,
\newblock {\it JHEP}, 07:049, 2002.

\bibitem{Ambjorn:1984yu}
J.~Ambj\o{}rn, P.~Olesen, and C.~Peterson,
\newblock {\it Nucl. Phys.}, B244:262, 1984.

\markright{Bibliography}

\bibitem{Majumdar:2002mr}
P. Majumdar,
\newblock {\it Nucl. Phys.}, B664:213, 2003.

\bibitem{Michael:1994ej}
C.~Michael and P.~W. Stephenson,
\newblock {\it Phys. Rev.}, D50:4634, 1994.

\bibitem{Teper:1998te}
M.~J. Teper,
\newblock {\it Phys. Rev.}, D59:014512, 1999.

\bibitem{Lucini:2004my}
B. Lucini, M. Teper, and U. Wenger,
\newblock Glueballs and k-strings in SU(N) gauge theories : calculations with
  improved operators,
\newblock hep-lat/0404008 .

\bibitem{deForcrand:1985cz}
P.~de~Forcrand, G.~Schierholz, H.~Schneider, and M.~Teper,
\newblock {\it Phys. Lett.}, 160B:137, 1985.

\bibitem{Kaczmarek:1999mm}
O. Kaczmarek, F. Karsch, E. Laermann, and M. L{\"u}tgemeier,
\newblock {\it Phys. Rev.}, D62:034021, 2000.

\bibitem{Kaczmarek:2003ph}
O. Kaczmarek, S. Ejiri, F. Karsch, E. Laermann, and F.  Zantow,
\newblock Heavy quark free energies and the renormalized Polyakov loop in full
  QCD,
\newblock hep-lat/0312015.

\bibitem{Bigi:1994em}
I.~I. Bigi, M.~A. Shifman, N.~G. Uraltsev, and A.~I. Vainshtein,
\newblock {\it Phys. Rev.}, D50:2234, 1994.

\bibitem{Beneke:1998rk}
M.~Beneke,
\newblock {\it Phys. Lett.}, B434:115, 1998.

\bibitem{Bali:2003jq}
G.~S. Bali and A. Pineda,
\newblock {\it Phys. Rev.}, D69:094001, 2004.

\bibitem{Appelquist:1977tw}
T. Appelquist, M. Dine, and I.~J. Muzinich,
\newblock {\it Phys. Lett.}, 69B:231, 1977.

\bibitem{Brambilla:1999qa}
N. Brambilla, A. Pineda, J. Soto, and A. Vairo,
\newblock {\it Phys. Rev.}, D60:091502, 1999.

\bibitem{Curci:1984wh}
G.~Curci, G.~Paffuti, and R.~Tripiccione,
\newblock {\it Nucl. Phys.}, B240:91, 1984.

\bibitem{Wohlert:1985hk}
R.~Wohlert, P.~Weisz, and Werner Wetzel,
\newblock {\it Nucl. Phys.}, B259:85, 1985.

\bibitem{Heller:1985hx}
U.~Heller and F.~Karsch,
\newblock {\it Nucl. Phys.}, B251:254, 1985.

\bibitem{Bali:2002wf}
G.~S. Bali and P. Boyle,
\newblock Perturbative Wilson loops with massive sea quarks on the lattice,
\newblock hep-lat/0210033.

\bibitem{ch1.NN.Stoks} V. Stoks, R. Timmermans and J. J. de Swart,
{\it Phys. Rev.},  C47:512, 1993, nucl-th/9211007 

\bibitem{ch1.QQpot.God.Is}
S. Godfrey and N. Isgur, {\it Phys. Rev.}, D32:189, 1985.

\bibitem{ch1.instpot} S. Chernyshev, M. A. Nowak, and I. Zahed, 
{\it Phys. Rev.},  D53:5176, 1996.

\bibitem{ch1.QQreg}
A. C. Mattingly and P. M. Stevenson, {\it Phys. Rev.}, D49:437, 1994.

\bibitem{ch1.QQdelreg} T. A. L\"{a}hde, C. J. Nyf\"{a}lt and D .O. Riska,
{\it Nucl. Phys.},  A674:141, 2000, hep-ph/9908485 


\bibitem{ch1:Bolder00} B. Bolder {\it  et al.}, {\it Phys. Rev.},  D63:074504,
 2001.


\bibitem{ch1.BrysNN} R. Bryan and B.L. Scott,  {\it Phys. Rev.}, 177:1435,
1969.

%july
\end{thebibliography}
%\input{chap1.bbl}
%july

\end{document}